\documentclass[12pt]{article}
\usepackage{amsmath,amsfonts, amssymb, braket, bbold}
\usepackage{tikz}
\usetikzlibrary{trees,er,snakes,shapes,mindmap}
\usepackage{titlesec}
\titleformat{\section}
 {\normalfont\fontfamily{put}\fontsize{12pt}{16pt}\bfseries\color{black}}
{\thesection}{1em}{}
\titleformat{\subsection}
 {\normalfont\fontfamily{put}\fontsize{12pt}{16pt}\bfseries\color{black}}
{\thesubsection}{1em}{}
\def \beq  {\begin{equation}}
\def \eeq  {\end{equation}}
\def \beqar {\begin{eqnarray}}
\def \eeqar {\end{eqnarray}}
\allowdisplaybreaks
\def\sqr#1#2{{\vcenter{\vbox{\hrule height.#2pt
\hbox{\vrule width.#2pt height#1pt \kern#1pt
\vrule width.#2pt}\hrule height.#2pt}}}}

\def\S {{\cal S}}
\def\la {{\langle}}
\def\ra {{\rangle}}
\def\vx {{\vec x}}

\def\vk {{\vec k}}
\def\vf {{\varphi}}

\def\bb {\bar{b}}

\def\bu {{\bar u}}

\def\bw{{\bar w}}

\def\bz {{\bar{z}}}
\def\vk {\vec{k}}
\def\vx {{\vec x}}

\def\bxi {{\bar \xi}}

\def\del {\partial}
\def\bdel{\bar{\partial}}
\def\a {\alpha}
\def\b {\beta}
\def\e {\epsilon}

\def\o {\omega}
\def\g {\gamma}

\def\C {{\cal C}}

\def\vf {{\varphi}}

\def\half{\textstyle{1\over 2}}

\mathchardef\mhyphen="2D
\begin{document}
\fontfamily{put}\fontsize{12pt}{16pt}\selectfont
\def \CMP {{Commun. Math. Phys.}}
\def \PRL {{Phys. Rev. Lett.}}
\def \PL {{Phys. Lett.}}
\def \NPBProc {{Nucl. Phys. B (Proc. Suppl.)}}
\def \NP {{Nucl. Phys.}}
\def \RMP {{Rev. Mod. Phys.}}
\def \JGP {{J. Geom. Phys.}}
\def \CQG {{Class. Quant. Grav.}}
\def \MPL {{Mod. Phys. Lett.}}
\def \IJMP {{ Int. J. Mod. Phys.}}
\def \JHEP {{JHEP}}
\def \PR {{Phys. Rev.}}
\def \JMP {{J. Math. Phys.}}
\def \GRG{{Gen. Rel. Grav.}}
\begin{titlepage}
\null\vspace{-62pt} \pagestyle{empty}
\begin{center}
\vspace{1.3truein} {\large\bfseries
Time-dependent Magnetic Fields and the Quantum Hall Effect}
\\
{\Large\bfseries ~}\\
\vskip .5in
{\Large\bfseries ~}\\
{\sc T.R. Govindarajan$^a$ and V.P. Nair$^b$}\\
\vskip .2in
{$^a$\sl The Institute of Mathematical Sciences, Chennai 600 113, 
India\\
Krea University, Sri City, 517646, CCSP, SGT University, Gurugram 122505,  India}\\
{$^b$\sl Physics Department,
City College of New York, CUNY\\
New York, NY 10031}\\
 \vskip .1in
\begin{tabular}{r l}
{\sl E-mail}:&\!\!\!{\fontfamily{cmtt}\fontsize{11pt}{15pt}
\selectfont vpnair@ccny.cuny.edu}\\
&\!\!\!{\fontfamily{cmtt}\fontsize{11pt}{15pt}
\selectfont trg@imsc.res.in, govindarajan.thupil@krea.edu.in}
\end{tabular}
\vskip .5in

\centerline{\large\bf Abstract}
\end{center}
Ermakov has shown how the solution to the classical
harmonic oscillator in one spatial dimension with general
time-dependent frequency can be reduced to the time-independent case and an associated nonlinear ordinary differential equation, an analysis which has been applied to the
Schr\"odinger equation as well. We extend this analysis
to the Landau problem of a charged particle in a uniform magnetic field in two
dimensions and construct the {\it {generalized}} Laughlin wave functions for the 
case when the magnetic field is time-dependent.
We also work out the dynamics of density fluctuations (the Girvin,MacDonald,Platzman or
GMP mode) and argue that it is possible to 
tune the frequency of the magnetic field to obtain a
compressible droplet of fermions.
We also analyze the dynamics of the edge modes of the droplet
for the integer Hall effect.

\end{titlepage}
\fontfamily{put}\fontsize{12pt}{16pt}\selectfont
\pagestyle{plain} \setcounter{page}{2}
\section{Introduction}
Ever since its discovery in the 1980s, 
quantum Hall effect has been a fascinating phenomenon, 
generating an enormous amount of research both on the 
experimental and theoretical aspects of the problem\cite{genQHE}.
An elaborate and successful theoretical structure based on wave functions, flux attachment procedures, conformal field theory, etc. has been built up by now \cite{fluxattach}.
We may also note that the Hall effect has been theoretically generalized to higher dimensions \cite{higherdimQHE1}-\cite{higherdimeQHE3}.  Some of the results or predictions obtained in this case can perhaps be accessible experimentally using the idea of synthetic dimensions \cite{syndim}.
The mathematical aspects of the quantum Hall effect, including connections to noncommutative geometry and von Neumann algebras, have also been
of great research interest \cite{NCgeom}.
For the standard quantum Hall effect in 2+1 dimensions, 
we may also note that a
Chern-Simons action, which includes the electromagnetic vector potential
as well as a statistical gauge field, has been shown to apply to the bulk dynamics of a droplet of electrons \cite{QHECS}.
For such a droplet, which is a state of finite volume with a boundary,
the edge dynamics, which is given by a chiral boson field, is important as well.
Among other features, it helps to cancel the gauge anomaly resulting
from the variation of the Chern-Simons bulk action.
The response function of the system to variations of 
the electromagnetic field, for example, the Hall conductivity,
is also directly obtained from the Chern-Simons action.
Even though a flat background metric is what is relevant for
almost all experimental situations, considerations of the Hall effect
on spaces with a nontrivial background metric and/or nontrivial topology
are useful \cite{{genQHE},{avron}}. They lead to the identification of new transport coefficients.
For instance, the term
proportional to the time-derivative of the metric in the
energy-momentum tensor of the system gives the Hall viscosity.
For this analysis, one is thus looking at a time-dependent
(albeit slowly varying) background
 metric. A natural question is then: What is the dynamics of a quantum Hall state if the magnetic field, which is used to define the Landau levels and
 the Hall state, is itself time-dependent?
 This is the subject we focus in the present paper. 
 Even though there is a large body of literature on the Hall effect, 
this particular question seems not to have received any attention.
 We do take note that adiabatic changes of the magnetic field have been 
considered before, see the review \cite{adiabaticB}.
 Also the case of time-dependent electrostatic or confining potentials
 has been analyzed in \cite{V(t)}.

Our approach to the problem of time-dependent magnetic fields starts with a 
simple observation. For the harmonic oscillator with a time-dependent 
frequency $\omega (t)$, the general solution of the time-dependent 
Schr\"odinger equation can be constructed provided one can solve
an auxiliary nonlinear equation known as the Ermakov equation
\cite{Ermakov}. This method is briefly reviewed in the next section.
Since the single-particle Landau problem (of a charged particle in a uniform 
magnetic field) is reducible to a harmonic oscillator, we can expect
that a similar strategy can be applied.
But the Landau problem of interest is in two spatial dimensions
and the degeneracy due to magnetic translations must be taken into account.
The holomorphicity of the lowest Landau level wave functions
is a key feature of the Laughlin-type many-particle wave functions.
Therefore, one would like to express the solution to the 
time-dependent Schr\"odinger equation in terms of coherent states 
of the oscillator with time-independent frequency.
This means that a simple generalization of the Ermakov method is needed, 
and this will be presented in section 3. We will also construct the 
Laughlin-type states for filing fraction $\nu = (1/2p+1)$.
The change in magnetic flux due to the time-dependence of the magnetic field
$B$ will create electric fields (by Faraday's law) and consequent currents.
We work out the expressions for the charge and current
densities as well.

But even more significantly, while for a constant magnetic field
a droplet of electrons behaves as an incompressible fluid,
the time-dependence of $B$ allows for compression and dilation 
since the magnetic length of a single-particle state can change with time.
(We consider keeping the filling fraction fixed as we change $B$.)
This will also drive the dynamics of the Girvin-MacDonald-Platzman (GMP)
mode \cite{GMP}. The relevant equations are worked out in section 4.
As an example, we also
consider a magnetic field of the form
$B (t) = B_0 + B_1\, \sin \Omega t$, where $B_0$
is the constant (time-independent) magnetic field 
and the second term is a time-dependent perturbation of it.
Such a magnetic field will drive the
GMP mode; it can then have frequencies $\omega_k \pm \Omega$ 
where $\omega_k$ is the unperturbed
frequency of the GMP mode.
These should be 
experimentally detectable, perhaps, in a suitable scattering process.
Also, for a suitable $\Omega$, one may even be able to eliminate the gap
of the GMP mode and obtain a transition to a compressible situation.
(This could be a compressible fluid or a crystal depending on the wave numbers for the energy profile of the GMP mode.)

A natural extension of our analysis would be to a droplet of fermions.
For a constant magnetic field, these are described by area-preserving 
diffeomorphisms of the droplet.
The time-dependence of $B$ will affect the edge dynamics
since compression and dilation are now additional modes.
For the case of the integer Hall effect, we work out
the generalization of the area-preserving diffeomorphsims
and the action for the edge modes in section 5.
The effect of allowing for compression and dilation or changes in the 
radius of a droplet as the magnetic field changes is that the equation
governing the edge modes becomes an integro-differential equation.

The analysis of the edge modes for the fractional QHE will be 
significantly more involved due to the fact that interparticle interactions
are a key to obtaining the required states.
With a time-dependent magnetic field, it
is not clear at this point how the interactions respond to
area-preserving diffeomorphisms. This issue, as well as the question of
solving the equations for the edge fluctuations, will be left to future
work.

We conclude with a short resume of the results in section 6.
There are two appendices giving technical details of the simplification of
the actions for
density fluctuations and edge modes.
\section{Ermakov method for the 1d-oscillator with time-dependent frequency}
We start by briefly recalling the salient features of the one-dimensional oscillator
with time-dependent frequency.
The Hamiltonian for this case is given by
\beq
H(t)~=~\frac{1}{2}\Bigl(p^2~+~\o^2(t)q^2\Bigr)
\label{Ham}
\eeq
The Schr\"odinger equation can thus be written out as
\beq
i {\del \Psi \over \del t} = {1\over 2}  \Bigl[ -\del_q^2 + \omega(t)^2\, q^2 \Bigr]\, \Psi
\label{erm1}
\eeq
In the Ermakov analysis, the solution to this equation is given by
\beqar
\Psi(t) &=& {1\over \sqrt{b} } e^{i \phi} \, \Psi_0 (\xi , 0)\nonumber\\
\phi &=& {{\dot b}\, q^2 \over 2 b} - E\, \tau, 
\hskip .2in \tau = \int_0^t dt' {1\over b^2(t)}
\label{erm2}
\eeqar
where $\xi = q/b$, with $b$ as a time-dependent scale factor. Here $\Psi_0 (\xi, 0)$ is the wave function obeying the
equation
\beq
{1\over 2}  \left[ - {\del^2 \over \del \xi^2}   + \omega(0)^2  \xi^2 \right]
\Psi_0 (\xi, 0) = E \, \Psi_0 (\xi, 0)
\label{erm3}
\eeq
Thus $\Psi_0$ is the wave function for a fixed frequency
$\omega(0)$ with energy $E$.
The allowed values for $E$ are $(n +{\half} )\omega (0)$ as usual.
By direct differentiation of $\Psi (t)$ as given in (\ref{erm2}), one can easily 
check that it is a solution of the Schr\"odinger equation (\ref{erm1}) if the scale factor $b$
obeys the equation
\beq
{\ddot b} + \omega^2(t) \, b = {\omega^2(0) \over b^3}
\label{erm4}
\eeq
This is the Ermakov equation. If we have a solution to this equation, we have an exact solution to the Schr\"odinger equation. Notice also that, from
(\ref{erm2})
\beq
\int dq\, \Psi^* \Psi = \int {dq \over b}\, \Psi_0^* (\xi, 0) \Psi_0 (\xi, 0)
=  \int {d\xi} \,\Psi_0^* (\xi, 0) \Psi_0 (\xi, 0)
\label{erm5}
\eeq
showing that if $\Psi_0$ is normalized, then so is $\Psi$.

The essence of the Ermakov method hinges on the
existence of an invariant for the oscillator even when $\omega$ is time-dependent.
Towards this, we can start with the general operator
\beq
I(t)~=~\frac{1}{2}\Bigl(\a(t)q^2~+~\b(t)p^2~+~\g(t) (q p +p q)\Bigr)
\label{intr2}
\eeq
The time-derivative of $I$ is given by
\beq
{\dot I} = {d I \over dt} = {\del I \over \del t} - i [I, H]
\label{intr3}
\eeq
where the commutator is evaluated by the standard rule
$[q, p] = i$.
Imposing time-independence of $I(t)$, i.e., ${\dot I} = 0$,
we get equations for the functions
$\a(t),\b(t),\g(t)$ \cite{LewisR}.
They can be solved in terms 
of $b(t)$ obeying (\ref{erm4}). The result for $I$ is then
\beq
I(t)~=~\frac{1}{2}\Bigl(\frac{q^2}{b^2}~+~(b~p~-~
\dot{b}q)^2\Bigr) 
\label{intr4}
\eeq
where, for brevity, we have set $\omega^2(0) = 1$.

It is also possible to construct the eigenvalues of the 
new invariant $I(t)$. Towards this, one
 can define new ladder 
operators $a$, $a^\dagger$ by
\beqar
a &=& \frac{1}{\sqrt{2}}\left(\frac{q}{b}~+~i(b p~-\dot{b}q)\right)\nonumber\\
a^\dagger &=& \frac{1}{\sqrt{2}}\left(\frac{q}{b}~-~i(b p~-\dot{b}q)\right)
\label{intr5}
\eeqar
It is easy to verify that these obey the expected 
commutation rule $\big[a~,a^\dagger\big]~=~1$.
The invariant $I$ is given in terms of these
operators by
\beq
I~=~\Bigl(a^\dagger~a~+~\frac{1}{2}\Bigr) 
\label{intr6}
\eeq
But we should remember our Hamiltonian (\ref{Ham}) 
is not diagonal in the
diagonal basis of $I(t)$.  

The key point here is the variable $\xi$ which is a scaled version
of $q$, with the scale factor obeying the Ermakov equation.
We can generalize the method to the two-dimensional problem 
of a charged particle in a magnetic field by identifying the
scaled version of the coordinates.
We turn to this now.
\section{The Ermakov equations for Landau levels}
For the Landau problem or the problem of a charged particle
in a magnetic field $B(t)$ with time dependence in two spatial dimensions, 
the Hamiltonian is given by

\beqar
H_0&=& {(p - e A )^2 \over 2 m} - \textrm{zero-point energy}\nonumber\\
&=& - {2 \over m} \left( {\del \over \del z} - \lambda \bz \right)
\left( {\del \over \del \bz} + \lambda z \right)
\label{erm6}
\eeqar
where $\lambda = {e B /4}$. In the second line we have introduced the
combinations
\begin{alignat}{3}
z &= x^1 + i x^2, \hskip .2in &\bz &= x^1 - i x^2\nonumber\\
{\del \over \del z}&= {\half} (\del_1 - i \del_2), \hskip .2in
&{\del \over \del \bz} &= {\half} (\del_1 + i \del_2)
\label{erm7}
\end{alignat}
The solution to the time-dependent Schr\"odinger equation with 
the Hamiltonian (\ref{erm6}) is then given by
\beq
\Psi  (z, \bz, t) = {1\over \sqrt{\bb b}}\, e^{i \Phi } \, \Psi_0 (\xi, \bxi, 0)
\label{erm8}
\eeq
where
$\xi = z/b$, ${\bar 
\xi} = \bz /{\bar b}$ and $\Psi_0(\xi, \bxi, 0)$ is an eigenstate of
$H$ at time $t =0$ with energy eigenvalue $E$.
Notice that we need a complex function $b$ in the present case.
It is convenient to write $b$ as
\beq
b = \sqrt{\rho} \, e^{i \theta}
\label{erm9}
\eeq
In terms of $\rho$ and $\theta$, the Ermakov equations are 
\beqar
{\ddot \rho} - {{\dot \rho}^2 \over 2 \rho} - { 8\over m^2} 
\left( {\lambda_0^2 \over \rho} - \lambda^2 \right) = 0&&
\label{erm10}\\
\theta = \int dt' {2 \over m} \left({ \lambda_0 \over \rho(t')} - \lambda (t') \right)&&
\label{erm11}
\eeqar
where $\lambda_0 = e B_0/4$.

The phase $\Phi$ in $\Psi$ in (\ref{erm8}) is given by
\beq
\Phi = {m {\dot \rho} \over 4 \rho } \bz z  - \theta (t) -
\int^t dt'  {E \over \rho(t')}
\label{erm12}
\eeq
Verification that (\ref{erm8}) does indeed satisfy the time-dependent Schr\"odinger equation is somewhat tedious but can be done in a straightforward way by 
direct differentiation. One can verify that  (\ref{erm8}) is the solution
 if $\Psi_0$
obeys the equation
\beq
 - {2 \over m} \left( {\del \over \del \xi} - \lambda_0 \bxi \right)
\left( {\del \over \del \bxi} + \lambda_0 \xi \right)
\Psi_0(\xi, \bxi, 0) = E \, \Psi_0 (\xi, \bxi, 0)
\label{erm13}
\eeq
The normalization of the wave function in (\ref{erm8}) is given by
\beq
\int d^2z\, \Psi^* \Psi = \int {d^2 z \over \rho} \,\Psi_0^* (\xi,\bxi,0) \,\Psi_0 (\xi, \bxi ,0)
= \int d^2\xi \, \Psi_0^* (\xi,\bxi,0)\, \Psi_0 (\xi, \bxi ,0) 
\label{erm14}
\eeq
We see that $\Psi$ is normalized if $\Psi_0$ is normalized.
In order to work out the time-dependence of $\Psi$ explicitly, 
we need to solve the Ermakov equations. In principle, one can solve
(\ref{erm10}) for $\rho(t)$ and then use it to obtain
$\theta$, $\Phi$ and $\Psi$. Admittedly, solving the nonlinear equation
(\ref{erm10}) is difficult in general. But the point for us is not
the explicit solution, but rather (\ref{erm8}) gives the general form
of the wave function and hence it can be used to construct
multiparticle wave functions to analyze density fluctuations and 
edge modes.

We will now focus on the lowest Landau level for which the solutions
to (\ref{erm13}) may be written as
\beq
\Psi_{0n} = e^{- \lambda_0 \bxi \xi} \, {\xi^n \over \sqrt{n!}}
=e^{ - \bz z /2 \kappa} \, {\xi^n \over \sqrt{n!}}
\label{erm15}
\eeq
where $\kappa = (\rho / 2 \lambda_0)$.
The Slater determinant of $\Psi$ from (\ref{erm8}) with the choice
 (\ref{erm15}) for $\Psi_0$ will give the wave function for the 
 $\nu = 1$ quantum Hall state. Similarly, for the
 $\nu = {1\over 2 p +1}$ state, we can write down the Laughlin wave function as
 \beq
 \Psi^{(2p+1)} (x_1, x_2, \cdots, x_N)
 = {\C_{2p+1} \over \rho^{N\over 2}} \exp\left(
\sum_i i \Phi_i - {\bz_iz_i \over 2 \kappa} \right)\,
 \prod_{i<j} (\xi_i - \xi_j )^{2p+1}
 \label{erm16}
 \eeq
 Since the phases cancel out in $\Psi^* \Psi$, we notice that
 \beq
 \int \prod_i dz_i d\bz_i \, \Psi^* \Psi = 
 \vert \C_{2p+1}\vert^2 \int \prod_i d\xi_i d\bxi_i ~
 e^{- \sum_i \lambda_0 \bxi_i \xi_i} \prod_{i<j } \vert \xi_i -\xi_j\vert^{2 (2p +1)}
 \label{erm17}
 \eeq
This shows that the normalization factor $\C_{2p+1}$ in (\ref{erm16})
will be independent of $\rho$ and other time-dependent factors.

It may be useful to collect the notational abbreviations we have introduced so far for ease of comparison with later sections:
\beq
\lambda = {e B \over 4}, \hskip .2in
\lambda_0 = {e B_0 \over 4}, \hskip .2in
\kappa = {\rho \over 2 \lambda_0} = {{\bar b} b \over 2 \lambda_0},
\hskip .2in \xi = {z\over b}
\label{erm17a}
\eeq
$\rho$ and $\theta$ obey the equations (\ref{erm10}), (\ref{erm11})
and $\Phi$ is given in (\ref{erm12}).

A few comments about the wave function (\ref{erm16})
are in order at this point.
First of all, notice also that if $B$ is independent of time, i.e., if $\lambda_0 = 
\lambda$, the solution to (\ref{erm10}, \ref{erm11}) is 
$\rho =1$, $\theta = 0$, i.e., $b = 1$, so that
$\xi = z$, ${\bar \xi} = \bz$. In this case (\ref{erm16}) reduces to the standard Laughlin-type state with $\nu = 1/(2 p+1)$.

We may also recall that there is no derivation of the Laughlin wave function from first principles. 
Rather its justification is based on numerical work, the fact that it
reproduces the observed fractional charge and statistics and that it can 
describe collective excitations. Thus ultimately the
correlation with experiments is what justifies it although there are
suggestive theoretical 
arguments using the composite boson or fermion picture.
While these justifications are not available for the case of the
time-dependent magnetic field, the Ermakov analysis shows that
the structure of the single-particle wave functions remains the same as for
the time-independent case, apart from a scaling of the coordinates
by a time-dependent factor and an overall phase.
Therefore, at least in an adiabatic sense, we could expect
the wave function (\ref{erm16}) to be applicable to the
$\nu = 1/(2 p+1)$ state with a time-dependent magnetic field.

Before moving on to considerations of excitations in the system, it is useful
to phrase these results in the language of field operators.
Introducing the standard fermion fields $\psi (x)$, $\psi^\dagger (x)$, we can write
\beq
H =  - {2 \over m} \int d^2x\, \psi^\dagger
\left( {\del_z} - \lambda \bz \right)
\left( {\del_\bz} + \lambda z \right) \psi
+ {\rm Interaction~terms}
\label{erm18}
\eeq
The interaction terms could include a Haldane pseudopotential 
to facilitate
obtaining the the Laughlin state as an eigenstate of $H$.
The Laughlin-type $\nu = 1/(2p+1)$ state can be expressed as
\beq
\ket{\nu} = \int \prod_i d^2x_i \,
\psi^\dagger (x_1)\, \psi^\dagger (x_2) \cdots \psi^\dagger (x_N)\,
\Psi^{(2p+1)} (x_1, x_2 \cdots, x_N) \, \ket{0}
\label{erm19}
\eeq
These field operators are in the Schr\"odinger picture and are 
independent of time. The time-dependence is carried by the wave functions
$\Psi^{(2p+1)}$. The state $\ket{\nu}$ obeys the equation
\beq
i {\del \over \del t} \ket{\nu} = H \ket{\nu}
\label{erm20}
\eeq
with $H$ as in (\ref{erm18}), by virtue of the fact that $\Psi$ obeys the time-dependent
Schr\"odinger equation. 
We also define a charge density $e J^0$ where
$J^0 = \psi^\dagger \psi$. The commutator of this with the Hamiltonian
gives
\beq
[H, \psi^\dagger \psi (x)] = i (\del {\bar J} + \bdel J )
\label{erm21}
\eeq
where
\beqar
{\bar J} &=&J_\bz = J^z = - i {2\over m} \, \psi^\dagger (\bdel + \lambda z) \psi
\nonumber\\
J&=& J_z = J^\bz = i {2\over m} \, (\del + \lambda \bz ) \psi^\dagger \, \psi
\label{erm22}
\eeqar
On the wave function (\ref{erm16}), $\bdel$ only acts on the exponents
since the rest of the wave function is holomorphic. It is therefore straightforward to evaluate the charge density and currents for the states $\ket{\nu}$.
We get
\beqar
\bra{\nu} J^0 \ket{\nu} &=& \bra{\nu} \psi^\dagger \psi \ket{\nu} = 
N \, {\cal F} (z,\bz)\nonumber\\
\bra{\nu} {\bar J} \ket{\nu} &=&N \, {\cal F} (z,\bz)\left[  - i {4 \lambda_0 \over m \rho} \alpha \right]\, z  = - i {2 \over m \kappa} \alpha\, z \bra{\nu} J^0\ket{\nu}\label{erm23}\\
\bra{\nu} {J} \ket{\nu} &=&N \, {\cal F} (z,\bz)\left[   i {4 \lambda_0 \over m \rho} \alpha^*  \right] \,\bz =   i {2 \over m \kappa} \alpha^*\, \bz \bra{\nu} J^0\ket{\nu}\nonumber
\eeqar
where
\beq
{\cal F} (z,\bz) ={\vert \C_{2p+1}\vert^2  \over \rho}
e^{- \bz z /2 \kappa} \int
\prod_{2}^N d^2\xi_i e^{- \sum_2^N \bxi \xi} 
\prod_{i\neq 1} \vert \xi - \xi_i\vert^{2 (2p +1)}
\prod_{i,j =2, i<j}^N  \vert \xi_i - \xi_j\vert^{2 (2p +1)}
\label{erm24}
\eeq
and
\beq
\alpha = \lambda \kappa - {1\over 2} + i {m {\dot \kappa }\over 4}
\label{erm25}
\eeq

We want to emphasize that the specific form of $H$ is actually not crucial for the expressions for the charge density and the currents. The expression for $J^0$
follows from 
\beq
J^0 = \int d^2z_2 \, d^2z_3 \cdots d^2z_N \, \Psi^{(p+1)*} \Psi^{(2p+1)}
\label{erm25a}
\eeq
Current conservation which is of the form, and which also follows
from (\ref{erm21}),
\beq
{\del \over \del t} \bra{\nu} J^0 \ket{\nu} = 
\del  \bra{\nu} {\bar J} \ket{\nu} + \bdel \bra{\nu} J \ket{\nu}
\label{erm26}
\eeq
can then be used to infer the expressions for $J$, ${\bar J}$
up to a term of zero divergence. Also, if the interaction terms in
(\ref{erm18}) involve products of the density, as for the pseudopotentials,
then gauging with respect to an external vector potential 
$A_i$ only modifies the
first term in $H$ and hence the operator expressions (\ref{erm22}) are obtained
by variation with respect to $A_i$.

If the magnetic field has no time-dependence, then
$\rho =1$ and $\alpha = 0$. The currents vanish in this case
as expected. The change of magnetic field with time induces an
electric field in the azimuthal direction, which, via the Hall conductivity,
gives radial currents as in (\ref{erm23}). 
This observation was made in the context of adiabatic changes
of $B$ in \cite{adiabaticB}.

To summarize this section, the single-particle wave functions are
given in (\ref{erm8}) with the Ermakov equations as in (\ref{erm10}), 
and (\ref{erm11}) and $\Psi_0$ obeying (\ref{erm13}).
The wave function for the $\nu= {1\over 2p +1}$ quantum Hall state is given
in (\ref{erm16}).
\section{Density fluctuations}
We now introduce a state with density fluctuations
by
\beq
\ket{f} =  J^0(f) \ket{\nu} =  \int \psi^\dagger \psi (x) f(x)\, \ket{\nu}
\label{df1}
\eeq
In the usual analysis (as in Feynman's work on liquid Helium 
\cite{Feyn1} and its use in the context of quantum Hall effect \cite{{adiabaticB},{GMP}}) one minimizes
the expectation value of the Hamiltonian for the state
$\ket{f}$ to determine $f$ and the spectrum of low-lying excitations
for the density fluctuations.
This strategy does not apply to the present case since we have a time-dependent problem. So we will write down an action for
the states in (\ref{df1}) and extremize it to obtain the equations of
motion.
This action is given by
\beq
\S [ f] = \int dt \left[ {i \over 2}  \bra{f} \, {\del \ket{f}\over \del t} 
- {i \over 2}{\del \bra{f} \over \del t} \, \ket{f} 
- \bra{f} H \ket{f} \right]
\label{df2}
\eeq
In working out various terms in this action, it should be kept in mind that
 $f$ and $\ket{\nu}$ in (\ref{df1}) are time-dependent.
The simplification of this action is straightforward 
and leads to the result
\beqar
\S[f] &=&  \int d^2x d^2y\, {i \over 2} \left[ f^*(x) {\del f(y) \over \del t} - 
{\del f^*(x) \over \del t} f(y) \right] S(x, y)\nonumber\\
&&- {2\over m} \int d^2x\, \del f^* \, \bdel f  \langle
\psi^\dagger \psi (x) \rangle
- {i {\dot \kappa}  \over 2 \kappa^2} 
\int d^2x d^2z \, f^*(x) f(z) ({\bar x} x - \bz z)  S(x, z)
\label{df11b}
\eeqar
where $S(x, y)$ is the two-point density correlation function
\beq
S(x, y) = \bra{\nu} \psi^\dagger \psi (x) \, \psi^\dagger \psi(y) \ket{\nu}
\label{df11c}
\eeq
Since there are many algebraic details involved, 
the steps from (\ref{df2}) to (\ref{df11b}) are
explained in Appendix A.

We can now take the variation of the action (\ref{df11b}) with respect to $f^*(x)$
 and obtain the equations of motion as
 \beqar
i \int d^2z\,  {\del f (z)  \over \del t} \, S(x,z) &=& 
 - {2 \langle \psi^\dagger \psi\rangle \over m} \bdel \del f(x)
 - {i \over 2} \int d^2z\, {\del S(x,z) \over \del t} \, f(z)\nonumber\\
 &&+ i {{\dot \kappa} \over 2 \kappa^2} 
 \int d^2z\, ({\bar x} x - \bz z) S(x, z) f(z) 
 \label{df13}
 \eeqar
We have taken the density $\langle \psi^\dagger \psi\rangle$ to be 
uniform for the unperturbed system, i.e., for the state $\ket{\nu}$.
(It is still a function of time.)
As a check on this result, notice that, if the magnetic field is constant
in time,
${\dot \kappa} = 0$ and $(\del S /\del t) = 0$.
In this case we can solve (\ref{df13}) with the ansatz
$f(x) = f_k e^{-i \omega_k t + i \vk \cdot \vx}$ to obtain 
$\omega_k = {k^2 \la \psi^\dagger \psi\ra / (2 m S(k))}$. This is the well-known result for the case of a time-independent magnetic field.
For the Laughlin states, $S(k) \sim k^2$, showing that there is an energy
gap for the density fluctuations, which is a signal of the 
incompressibility of the electron droplet.

Equation (\ref{df13}) gives the dynamics of the density fluctuations
in the present case where the magnetic field can be time-dependent.
To proceed further one needs to know the time-dependence and the resulting solution of the Ermakov equations (\ref{erm10}), (\ref{erm11}).
To get a sense of the dynamics, we will now consider
the analysis of the equation of motion
(\ref{df13}) for a magnetic field of the form
$B = B_0 + B_1 \sin \Omega t$
where $B_0$
which is constant in time is the dominant term and 
$B_1 \sin  \Omega t$ gives a harmonic perturbation.
Assuming the magnitude of $B_1$ is small compared to $B_0$
we can expand various quantities of interest to first order as
\beqar
f &=& f_0 + f_1 + \cdots\nonumber\\
S&=& S_0 + S_1 + \cdots\nonumber\\
\rho &=& 1 + \rho_1 + \cdots
\label{df14}\\
\la \psi^\dagger \psi\ra&=& \la \psi^\dagger \psi\ra_0 + \la \psi^\dagger \psi\ra_1
+ \cdots,\nonumber
\eeqar
equation (\ref{df13}) can be approximated by
\beqar
i \int_z {\del f_0(z) \over \del t} S_0(x,z)  &=&  - {2 \langle \psi^\dagger \psi\rangle_0 \over m} \bdel \del f_0(x)\label{df15}\\
i \int_z {\del f_1(z) \over \del t} S_0(x,z)
+ {2 \langle \psi^\dagger \psi\rangle_0 \over m} \bdel \del f_1(x)
&=&\biggl[ -i \int_z {\del f_0(z) \over \del t} S_1(x,z) 
- {i \over 2} \int_z {\del S_1(x,z) \over \del t} f_0 (z)\nonumber\\
&&~~- {2 \la \psi^\dagger \psi\ra_1 \over m} \bdel \del f_0(x)
\nonumber\\
&&~~+ i {{\dot \kappa} \over 2 \kappa^2} \int_z ({\bar x} x - z\bz ) S_0(x,z) f_0 (z)
\biggr]
\label{df16}
\eeqar
It is useful to write $B_1 \sin \Omega t = -i (B_1/2)(e^{i \Omega t}
- e^{ -i \Omega t})$ since (\ref{df16}) is linear in the perturbation and so
each exponential can be treated separately.
When the magnetic field is varying with frequency $\Omega$,
the induced electric field has the same frequency. As a result, the current
which is proportional to this electric field, the charge density $\la \psi^\dagger \psi\ra_1$,
which is related to the current by conservation law will have the same frequency. Further, we can simplify the Ermakov equation
(\ref{erm10}) as
\beq
{\ddot \rho_1} + {8 \lambda_0\over m^2} \left( \lambda_0 \rho_1 + 2 \lambda_1 \right) \approx 0
\label{df17}
\eeq
This shows that $ {{\dot \kappa} \over 2 \kappa^2}$ in (\ref{df16}) will also have the same frequency.
Since we have $f_0$ on the right hand side of (\ref{df16}), the
terms of the right have a factor $e^{ i (\omega_k  \pm \Omega ) t}$.
This will be the driving frequency for $f_1$, which will thus carry a factor
$e^{ i (\omega_k  \pm \Omega ) t}$ as well.
In the same way that the GMP modes and the energies $\omega_k$
can be seen in Raman scattering, it should be possible to observe these
modified energies as well.
The energies $\omega_k$ have a gap, the lowest value being the
magnetoroton gap. So in addition to the observation of
the driven frequencies, by tuning $\Omega$ it should be possible to
get $\omega_k - \Omega = 0$ at the magnetoroton value. This would mean that 
the magnetoroton, and hence the density fluctuations, can become gapless 
at this point,
indicating a transition to a compressible fluid, with compression modes of
nonzero wave vector $k$.

To summarize, in this section, we have considered density fluctuations
or the GMP modes of the quantum Hall droplet in the bulk and obtained the equations of motion governing them (\ref{df16}) when the magnetic field itself can be time-dependent.
For small oscillating magnetic field around a constant value,
we obtain a frequency shift for these modes of the form
$\omega_k \pm \Omega$. 
This should be an observable effect; a tuning of the frequency $\Omega$ can lead to a compressible droplet.
\section{Edge dynamics for the QH droplet}
We now turn to the dynamics of edge states
of a droplet of $N$ fermions with some confining potential
$V$. 
For a constant time-independent magnetic field, the droplet is
incompressible and the edge excitations, therefore, correspond to
area-preserving diffeomorphisms.
For these transformations and their quantum counterparts, the so-called
$W_\infty$ algebras, see
\cite{area-pd1}, \cite{area-pd2}, as well as various reviews cited.
In the present case with time-varying magnetic field, we should expect similar 
edge excitations, but their dynamics will be modified by the time-dependence of the magnetic field.
This modified dynamics is what we will work out in this section.
The problem is complicated by the fact that there is explicit time-dependence
in various expressions.

We begin by considering area-preserving diffeomorphisms in the classical case. The area element is of the form
\beq
\omega = i \, dz\wedge d\bz
\label{e1}
\eeq
Consider the transformation $z \rightarrow z+ \e (z, \bz )$,
$\bz \rightarrow \bz + {\bar \e} (z, \bz )$. The change in the area element is given by
\beq
\delta \omega =
 i \left(  {\del \e \over \del z} + {\del {\bar \e} \over \del \bz} 
\right) dz\wedge d\bz
\label{e2}
\eeq
Setting this to zero, we find that the changes in coordinates are of the form
\beq
 \e = i{\del f \over \del \bz}, \hskip .2in {\bar \e} = - i {\del f \over \del z}
 \label{e3}
 \eeq
where $f$ is real. On a function $h$, this change can be obtained as
\beqar
h &\rightarrow& h + \e {\del h \over \del z} + {\bar \e} {\del h \over \del \bz}
= h + i \left( {\del f \over \del \bz } {\del h \over \del z} - {\del f \over \del z}
{\del h \over \del \bz }\right)\nonumber\\
&=&h +  \{ f, h\}
\label{e4}
\eeqar
where we define a Poisson bracket
\beq
\{ f, h \} = i \left( {\del f \over \del \bz } {\del h \over \del z} - {\del f \over \del z}
{\del h \over \del \bz }\right)
\label{e5}
\eeq

We will now identify the operator in the quantum theory which carries out the same transformation. Since the change in $h$ is given by the Poisson bracket
with $f$, we should expect that
the operator corresponding to $f$ should play this role in the quantum theory.
To show that this is indeed the case, consider the operator versions of
$z$ and $\bz$.
The wave functions of interest for us are of the form
\beq
\Psi (z, \bz, t )= {1\over \rho} e^{i \Phi} e^{-\bz z /2\kappa} h(z)
\label{e6}
\eeq
where $h(z)$ is a holomorphic function. (It is equal to $z^n/(b^n \sqrt{n!})$
for the wave functions shown in (\ref{erm15}).)
The phase $\Phi$ is given in (\ref{erm12}) and $\kappa = \rho/ (2 \lambda_0)$.
Eventually we are interested in the multiparticle wave functions, but they also have
the form of (\ref{e6}) with the exponential factor multiplying holomorphic functions.

Let us first look at the case where we do not have the phase factor
$e^{i \Phi}$.
The rest of the wave function is a coherent state with $K = \bz z/\kappa $ as the K\"ahler potential.
This suggests that we can consider the operators $Z$ and ${\bar Z}$ 
as defined by the quantization of the 
symplectic structure
\beq
\Omega = i\, {\del \bdel K } = {1\over \kappa} \, i dz \wedge d\bz =
{1 \over \kappa } \, \omega
\label{e8}
\eeq
Standard quantization then yields the result
\beqar
Z &=& z - \kappa D_\bz , \hskip .2in
D_\bz = \del_\bz + {z \over 2 \kappa}\nonumber\\
{\bar Z}&=&\bz + \kappa D_z , \hskip .2in
D_z = \del_z - { \bz\over 2 \kappa}
\label{e9}
\eeqar
The wave functions, again apart from $e^{i\Phi}$, obeys the polarization
condition
\beq
D_\bz \Psi = 0
\label{e10}
\eeq
Thus the quantum operators in (\ref{e9}) reduce to
\beq
Z = z, \hskip .3in {\bar Z} = \bz + \kappa D_z
\label{e11}
\eeq
The inclusion of the phase factor is straightforward.
Notice that the matrix element of an operator ${\hat A}$ can be written as
\beq
\bra{1}{\hat A} \ket{2} =
\int d^2z \, {\overline{h_1(z)}}\, e^{- \lambda_0 \bz z /( {\bar b} b)} 
\left(e^ {- i \Phi }\, {\hat A}\, 
e^ {i \Phi }\right) e^{- \lambda_0 \bz z /( {\bar b} b)}\, h_2(z)
\label{e12}
\eeq
Thus operators of the form 
$e^ {- i \Phi }\, {\hat A}\, 
e^ {i \Phi }$ act on the wave functions without the phase factor.
Interpreting (\ref{e11}) as applying to such combinations, we can write the
true form of the operators $Z$ and ${\bar Z}$ (acting on the wave functions
in (\ref{e6}) with the phase factor $e^{i \Phi}$) as
\beqar
Z\, \Psi &=&  e^{i \Phi} \left( z - \kappa D_\bz\right) e^{-i \Phi}\, \Psi
=  z \, \Psi\nonumber\\
{\bar Z}\, \Psi &=&e^{i \Phi} \left( \bz + \kappa D_z\right) e^{-i \Phi} \, \Psi
= e^{i \Phi} e^{- \bz z/2\kappa}~
\kappa {\del \over \del z} \, h(z)
\label{e13}
\eeqar
The modified polarization condition is
\beq
D_\bz \, e^{- i \Phi} \Psi = 0,
\label{e14}
\eeq
which gives wave functions of the same form as in 
(\ref{e6}), namely,
$e^{i \Phi} e^{-  \bz z/2 \kappa}\, h(z)$.
It is also easy to check that ${\bar Z}$ as given in
(\ref{e13}) is the adjoint of $Z$. This means that
${\bar Z}$ acting to the left gives $\bz$.
Also, from (\ref{e13}), the basic commutation rule
is given by
\beq
[{\bar Z}, Z ] = \kappa
\label{e15}
\eeq
Since $Z$ gives $z$ and ${\bar Z}$ via its action to the left gives
$\bz$, we can take $Z, {\bar Z}$ as the quantum operators corresponding to
$z, \bz$.

We can now establish a quantization rule and operator products.
We first write an operator ${\hat F}$ which is some
ordered function of ${\bar Z}$, $Z$ as
\beq
{\hat F} ({\bar Z} , Z ) = e^{{\bar Z} \del_{\bar w} } e^{Z \del_w} \, F({\bar w}, w)\Big\vert_{w, {\bar w} = 0}
\label{e16}
\eeq
We can then write 
\beqar
\int \Psi^* \, {\hat F}({\bar Z}, Z) \, \Psi &=& 
\int {\overline h(z)}\, e^{- \bz z/\kappa}  e^{ \kappa \del_z \del_\bw } 
F(\bw , z) \Big\vert_{\bw =0}\, h(z) \nonumber\\
&=& \int {\overline h(z)}\, e^{- \bz z/\kappa} 
F(\bz , z) \, h(z)
\label{e16a}
\eeqar
Notice that in (\ref{e16}) we have chosen an ordering for the operator where
${\bar Z}$'s are to the left. The use of (\ref{e13}) and
an integration by parts then leads to
(\ref{e16a}).
Thus the operator ${\hat F}$ in the integral with $\Psi^*$ and $\Psi$ can be replaced by $F(\bz, z)$.

If we consider the product of two operators ${\hat F}$ and ${\hat G}$, we get
\beqar
{\hat F} \, {\hat G} &=& e^{{\bar Z} \del_\bw } e^{Z \del_w} \, F(\bw, w)\,
e^{{\bar Z} \del_\bu } e^{Z \del_u} \,  G(\bu, u)\Big\vert_{w, \bw, u, \bu = 0}
\nonumber\\
&=& e^{{\bar Z} \del_\bw } e^{Z \del_w} \,
e^{{\bar Z} \del_\bu } e^{Z \del_u} \, F(\bw, w)\, G(\bu, u)\Big\vert_{w, \bw, u, \bu = 0}
\nonumber\\
&=& e^{{\bar Z} \del_\bw }e^{{\bar Z} \del_\bu }  e^{Z \del_w} \,
 e^{Z \del_u} \, e^{- \kappa \del_w \del_\bu }\,F(\bw, w)\, G(\bu, u)\Big\vert_{w, \bw, u, \bu = 0}\nonumber\\
 &=&e^{{\bar Z} \del_\bw }  e^{Z \del_w} \, (F*G)(\bw, w)\Big\vert_{w, \bw = 0}
 \label{e17}\\
(F*G)(\bw, w) &=&e^{- \kappa\del_w \del_\bu }\,F(\bw, w)\, G(\bu, u)\Big\vert_{u=w, \bu = \bw}\nonumber\\
&=& F (\bw, w) \, G(\bw, w) - \kappa \, {\del F \over \del w} 
{\del G \over \del\bw} + \cdots
\label{e18}
\eeqar
The last equation defines the star-product of the two functions
$F$ and $G$.
This result also shows that
\beq
\Psi^* ({\hat F} {\hat G}) \Psi = \Psi^*\, (F*G)(\bz, z) \, \Psi
\label{e19}
\eeq
From this, we also see that
\beq
\Psi^* \, e^{i {\hat F}} {\hat G}\, e^{-i {\hat F}} \, \Psi =
\Psi^* \, \left[ G(\bz, z)  + i \kappa \left( {\del F \over \del \bz } {\del G\over \del z} - {\del G \over \del \bz } {\del F\over \del z} \right) + \cdots\right]
\Psi
\label{e20}
\eeq
Comparing with (\ref{e5}), we see that $G$ undergoes an area-preserving diffeomorphism 
characterized by the function $f = \kappa F$.
We conclude that the deformation of the wave functions by an area-preserving diffeomorphism is given by $\Psi \rightarrow U\, \Psi$
with $U = e^{- i {\hat F}}$.
In other words, the state $e^{- i {\hat F}} \Psi$ gives the quantum version of the area-preserving diffeomorphism of $\Psi$.
For different choices of ${\hat F}$, $e^{- i {\hat F}} \Psi$ will describe the 
corresponding differently deformed states of the
droplet.

Our next step is the construction of the action corresponding to such
transformations.
A useful conceptual device for this is to
think of the droplet as a single composite particle with a field operator 
$\Sigma^\dagger$ for creating it and $\Sigma$ for destroying it. The action for $\Sigma$ is of the form
\beq
S = \int \left[ \Sigma^\dagger i {\del \Sigma \over \del t} - \Sigma^\dagger H \, \Sigma \right]
\label{e21}
\eeq
The operator $\Sigma$ has the form
\beq
\Sigma = \sum_\alpha A_\alpha \, \Psi_\alpha + \cdots
\label{e22}
\eeq
where $\Psi_\a$ are wave functions for different states of the droplet, including various 
area-preserving deformations of it, and the ellipsis denotes higher states, higher Landau levels, etc. $A_\alpha$ is the annihilation operator for
the state labeled by $\alpha$.
The starting state is a specific droplet configuration,
say, $\Psi_0 = \Psi^{(2p+1)}$. The path integral for its evolution is given by
\beq
\braket{f|i} = \int [d\Sigma] \, e^{i \S (\Sigma)}
\label{e23}
\eeq
The initial state is such that $\Sigma\ket{i} = \Sigma A_0^\dagger \ket{0} =
\Psi_0 \ket{0}$ and the final state
has $\bra{f} \Sigma = \bra{0} \Psi^*_f$.
In the approximation of neglecting higher states, we consider only  those paths or intermediate configurations in (\ref{e23}) which correspond to area-preserving
diffeomorphisms.
Thus the action reduces to
\beq
\S = \int \left[ \Psi^*  U^\dagger\, i {\del \over \del t} (U \Psi) 
- \Psi^* U^\dagger H' \, U \Psi \right]
\label{e24}
\eeq
where $U$ is of the form $e^{- i {\hat F}}$ as we have already
discussed, but now with all particles included, ${\hat F} = \sum_i {\hat F}({\bar Z}_i, Z_i )$.
We have dropped the subscript and superscript on $\Psi$ for brevity;
here
$\Psi = \Psi^{(2p+1)}$ as in 
(\ref{erm16}); it is also time-dependent
because of the time-dependence of $b$, ${\bar b}$, $\Phi$, etc.
Further the Hamiltonian $H'$ now includes the confining potential
$V$ as well; it is of the form $H' = H +V$, with $H$ as given in (\ref{erm18}).
Two-point correlations are important
for the unitary transformations
of the interparticle interaction terms leading to significant
complications in simplifying (\ref{e24}). We have not yet
completed these calculations, so, in the following we will confine 
our analysis to the integer Hall effect.
In this $H= H_0$, as given by the first term in (\ref{erm18}) or
equivalently by (\ref{erm6}).

The action for edge dynamics can now be obtained by straightforward simplification
of this using the star-products as in (\ref{e18}).
The first step is to write (\ref{e24}) in terms of derivatives of
${\hat F}$.
We take the  wave functions $\Psi$ to obey
\beq
i {\del \over \del t} \Psi = H'\, \Psi
\label{e27}
\eeq
Expanding $U = e^{-i {\hat F}}$ and using (\ref{e27}),
\beqar
\int  \Psi^* \, U^\dagger\, i {\del \over \del t} (U \Psi) 
&=& \int \Psi^* \left[ i\,U^\dagger {\del U \over \del t}  + H' \right] \Psi
\nonumber\\
&=& \int \Psi^* \left[ {\del {\hat F} \over \del t} + {i\over 2} \left( {\hat F} {\del {\hat F}\over \del t} - {\del {\hat F}\over \del t}  {\hat F}\right)
+\cdots + H'\right] \Psi
\label{e28}
\eeqar
Expanding the second term in $S$ to quadratic order in ${\hat F}$,
we get
\beq
\int \Psi^* \, U^\dagger H' U \, \Psi
= \int \Psi^* \left[ H' + i [{\hat F}, H'] - {1\over 2} [{\hat F}, [ {\hat F}, H']]
+ \cdots \right] \Psi
\label{e29}
\eeq
Notice that we can also use
\beq
{\del \over \del t} ( \Psi^* {\hat F} \Psi )
= \Psi^* \left[ {\del {\hat F} \over \del t} - i [ {\hat F}, H'] 
\right] \Psi
\label{e30}
\eeq
Thus, apart from a total derivative in $t$ (which integrates to an irrelevant
surface term at the initial and final time-slices and hence is a canonical
transformation), the action can be written as
\beq
\S \approx \int \Psi^* \left[ {i \over 2} ( {\hat F} {\dot{\hat F}} -
{\dot{\hat F}} {\hat F} ) + {1\over 2} [{\hat F}, [{\hat F}, H']]  + \cdots
\right] \Psi
\label{e31}
\eeq
The operators $Z$ and ${\bar Z}$ are time-dependent in general, so
the further simplification of this action will involve a few nontrivial technical steps. This is given in some detail in the Appendix B.
As shown there, the action (\ref{e31}) for a spherical droplet of radius $R$
simplifes as
\beqar
\S &=&- {1\over 4\pi} \int dt \oint d \theta  
\left[ {d F \over d t} {\del F \over \del \theta}
+ \left( \kappa {\del V \over \del r^2} + \beta\right) {\del F \over \del \theta} {\del F \over \del \theta}\right]\nonumber\\
&&
+ \int_D dt d^2x  {\alpha^* \alpha \over 2\pi \kappa m}
(\nabla F)^2
\label{e91a}
\eeqar  
where the subscript $D$ on the integral indicates that 
the region of integration is a disc of radius $R$ and $\beta$ is given by
\beq
\beta = {2 \over m} \left[ \lambda^2 \left( \kappa - {1\over 2 \lambda_0}
 \right) - \left( \lambda - {1\over 2 \kappa}\right) \right] 
 \label{e91b}
\eeq
The first two terms in (\ref{e91a}) give the action for a chiral bosonic mode.
It reduces to the standard result when the magnetic field is independent of
time, so that $\alpha, \, \beta = 0$. We can identify $v = \kappa (\del V/\del r^2)$ evaluated at the edge as the velocity, with energy $\sim v k$
for modes of the form $e^{i k \theta}$.
Our derivation of this result, via the use of star-products 
to simplify the action of the unitary transformation, is along the lines of
\cite{KN2}, modified to take account of the time-dependence of $B$.

It is fairly easy to write down the equations of motion. The
bulk equation of motion following from the last term in (\ref{e91a}) is
\beq
\nabla^2 F = 0
\label{e92}
\eeq
The equations for the edge dynamics is obtained as
\beq
{1\over 2\pi} \left[{d \over d t} \left( {\del F \over \del \theta}\right)
+ \left( \kappa {\del V \over \del r^2} +\beta\right) \, {\del^2 F \over \del \theta^2}
\right] + { \alpha^* \alpha  R \over \pi \kappa m}  \left({\del F \over \del r}\right)_{r=R}
= 0
\label{e93}
\eeq
The bulk equation (\ref{e92}) is a constraint, with no time-derivatives.
We can solve it in terms of the boundary value
of $F = F(t, R, \theta)$. If $G(x', x)$ is the Green's function for the Laplacian
obeying Dirichlet boundary conditions at $r = R$, Green's theorem tells
us that
\beq
F(x) = - \oint F(t, R, \theta' )\, \del_{r'} G(x', x)
\label{e94}
\eeq
This also gives
\beq
\del_r F = - \oint F(t, R, \theta' )\, M(\theta', \theta), \hskip .2in
M(\theta', \theta ) = \del_r \del_{r'} G(x', x)\big\vert_{r=r'=R}
\label{e95}
\eeq
The equation of motion (\ref{e93}) thus becomes the integro-differential equation
\beqar
{\del \over \del t} \left( {\del F \over \del \theta}\right)_R
- {\dot R} {\del \over \del \theta} \oint F(\theta' )\, M(\theta', \theta)
+ \left( \kappa {\del V \over \del r^2} +\beta\right) \, {\del^2 F \over \del \theta^2}&&\nonumber\\
 - {2 \alpha^* \alpha \over \kappa m} R \oint F(\theta' )\, M(\theta', \theta) &= 0&
 \label{e96}
\eeqar
where the ${\dot R}$ dependent term arises from writing
\beq
{d \over  dt} \left( {\del F \over \del \theta}\right)
= {d \over  dt} \left( {\del F \over \del \theta}\right)_R
+ {\dot R}  {\del \over \del \theta} \left( {\del F \over \del r}\right)_R
\label{e96a}
\eeq
Equation (\ref{e96})
is a closed equation for $F(t, R, \theta)$. Once we solve it, the value
of $F(t, R, \theta)$ can be used in (\ref{e94}) to obtain
$F$ in the bulk or the interior of the disc $D$.
The extra terms arising from the time-dependence of the magnetic field
involve the kernel $M(\theta', \theta)$, which, we may note, is the
Dirichlet-to-Neumann operator for the Laplacian
on the spatial manifold.

Needless to say, $\alpha$, $\beta$ and ${\dot R}$ are zero for the
case of the time-independent magnetic field, so
(\ref{e96}) reduces to the standard chiral boson dynamics of the edge
mode as expected.
We may also note that the choice of a boundary condition
$\del_r F = 0$ is not meaningful. Since
\beq
\int_D (\nabla F)^2 = \int_D F (-\nabla^2 F) + \oint F \del_r F
\label{e97}
\eeq
we see that the choice $\del_r F = 0$ on the boundary implies that
$\nabla F = 0$, given the bulk equation of motion (\ref{e92}).
This implies that $F$ is a constant in the disc and the value on the edge must also be the same constant by continuity.
So the choice of $\del_r F = 0$ as a boundary condition will not give any nontrivial
solutions.

In (\ref{e96}) we have set up the
framework for analyzing the dynamics of the edge modes. 
However we have not yet found any solution.
Solving it is a very involved problem and will be left
to future research.

\section{Conclusion}

In this paper, we have taken the first steps towards analyzing the quantum Hall effect when the magnetic field generating the Landau levels is time-dependent.
While the use of such fields is certainly physically viable, but
the theoretical analysis would seem to be quite difficult, at least at first glance.
Our key observation is that the Ermakov method can be adapted to this
problem. The problem of the harmonic oscillator with time-dependent frequency can be mapped to the oscillator problem with time-independent
frequency, with a scaling of coordinates. The scale factor is time-dependent and obeys a nonlinear equation. Originally formulated in the classical case, the method has been generalized to quantum mechanics \cite{Ermakov}, where an additional phase factor is also present.
We use the same method for the Landau problem, with a slight
generalization needed since we are in two dimensions.
The form of the single-particle wave functions remains the same as for the time-independent case, but with a time-dependent scale factor for the coordinates and an extra phase factor.
Therefore it is possible to write down the generalized form of the multiparticle wave functions. We obtain the generalization of
the $\nu = 1/(2p +1)$ Laughlin state, this is shown in (\ref{erm16}).

We also consider density fluctuations of a droplet of fermions generalizing the GMP analysis to the time-dependent case. This is worked out in section 4.
Considering small time-dependent perturbations around a time-independent
magnetic field, we find that the GMP mode acquires a frequency shift, presumably observable via Raman scattering.
More interestingly, it may be possible to tune the frequency of the perturbation to eliminate the energy gap of the GMP mode, indicating a transition to a
compressible situation.

We also considered how to set up the dynamics of the edge modes for a droplet of fermions for the integer Hall effect.
If the magnetic field is independent of time, the edge modes are the quantum version of area preserving diffeomorphisms of the droplet. 
Strictly speaking, it is the canonical structure relevant for the lowest Landau
level (or any fixed Landau level) that is preserved by the edge excitations.
This is proportional to the geometric area if the magnetic field is fixed.
But with a time-dependent magnetic field, the proportionality factor is
time-dependent and hence the connection to the geometric area is not straightforward.
This is the key conceptual complication regarding the edge modes.
In section 5, we showed how the analog of the area preserving diffeomorphisms can still be formulated.
We also construct the action governing the edge modes. This is
also a slight generalization of the usual action for the edge modes
(which is basically a coadjoint orbit action). 
An interesting point regarding the equations for the edge modes is the appearance of the Dirichlet-to-Neumann operator on the edge.
Unfortunately, the equation is still fairly complicated and we have 
not yet been able to find any useful solution.

As mentioned in the Introduction, the case of edge modes for the fractional case, when the magnetic field is time-dependent, is 
significantly more involved due to the key role of the interparticle
interactions. We have no definite results to report at this point, 
this matter is still under investigation.

\bigskip

VPN thanks Hans Hansson for discussions, suggestions for improving the
presentation and for bringing references
\cite{V(t)} to our attention.
We also thank Ajit Balram and Dimitra Karabali for discussions.

This work was supported in part by the U.S. National 
Science Foundation Grant No. PHY-2412479.

\section*{Appendix A: The action for density fluctuations}
\def\theequation{A\arabic{equation}}
\setcounter{equation}{0}
Here we carry out the simplification of the action
\beq
\S [ f] = \int dt \left[ {i \over 2}  \bra{f} \, {\del \ket{f}\over \del t} 
- {i \over 2}{\del \bra{f} \over \del t} \, \ket{f} 
- \bra{f} H \ket{f} \right]
\label{df2a}
\eeq
which is the cation (\ref{df2}) in text.
The kinetic term
can be evaluated in a straightforward way to get
\beqar
 {i \over 2}  \bra{f} \, {\del \ket{f}\over \del t} 
- {i \over 2}{\del \bra{f} \over \del t} \, \ket{f} 
&=& \int d^2x d^2y\, {i \over 2} \left[ f^*(x) {\del f(y) \over \del t} - 
{\del f^*(x) \over \del t} f(y) \right] S(x, y)\nonumber\\
&& + {1\over 2} \bra{\nu} \left[J^0(f^*) J^0(f) \, H 
+ H\, J^0(f^*) J^0(f) \right] \ket{\nu}
\label{df3}
\eeqar
where $S(x, y)$ is the two-point density correlation function
\beq
S(x, y) = \bra{\nu} \psi^\dagger \psi (x) \, \psi^\dagger \psi(y) \ket{\nu}
\label{df4}
\eeq
and the terms involving $H$ arise from the time-dependence of $\ket{f}$.

We can evaluate $\bra{f} H \ket{f}$ by moving $H$ to the right to act
on $\ket{\nu}$ or to the left to act on $\bra{\nu}$.
By moving $H$ to the right and using (\ref{erm21}), we find
\beqar
\bra{f} H \ket{f} &=& \int d^2x\, f(x)\bra{f} [H, \psi^\dagger \psi (x) \ket{\nu}
+ \bra{f} J^0(f) H \ket{\nu}\nonumber\\
&=& \int d^2x\, \left[ - i \del f \bra{f} {\bar J} \ket{\nu} 
- i \bdel f \bra{f} J \ket{\nu} \right] +  \bra{f} J^0(f) H \ket{\nu}
\label{df5}
\eeqar
For the commutator $[H, \psi^\dagger \psi]$ in this equation, we are basically using the conservation equation for the current, not a specific form of
$H$.
By using $\ket{\nu}$ as in (\ref{erm19}) and
 ${\bar J} $ from (\ref{erm22}), we find
\beqar
\int d^2z \,\left[ - i\del f(z) \bra{f} {\bar J}(z) \ket{\nu}
\right] &=& - {2\alpha \over m \kappa} \int d^2x d^2z\,
f^*(x) z\del f(z)  \, \bra{\nu} J^0(x) J^0(z) \ket{\nu} \nonumber\\
&=& - {2\alpha\over m \kappa} \int d^2x d^2z\,
f^*(x)\,  z\del_z f(z)  \, S(x, z)
\label{df6}
\eeqar
For $\bra{f} J \ket{\nu}$, since $J$ has $(\del + \lambda \bz )\psi^\dagger$
to the left, it is easier to move $J$ to the left of $J^0(f^*)$ to evaluate it.
This gives
\beqar
\int d^2z \,\left[ - i\bdel f(z) \bra{f} {J}(z) \ket{\nu}
\right] &=&{2\over m} \int d^2x\, \del f^* \, \bdel f  \bra{\nu} 
\psi^\dagger \psi (x) \ket{\nu} - i \int \bdel f \bar{\nu} J \, J^0(f^*) \ket{\nu}
\nonumber\\
&=&{2\over m} \int d^2x\, \del f^* \, \bdel f  \langle
\psi^\dagger \psi (x) \rangle\nonumber\\
&& + {2 \alpha^* \over m \kappa} 
\int d^2x d^2z  f^*(x) \,\bz \bdel_\bz f (z) \bra{\nu} 
J^0(z) J^0(x)\ket{\nu}\nonumber\\
&=&{2\over m} \int d^2x\, \del f^* \, \bdel f  \langle
\psi^\dagger \psi (x) \rangle\nonumber\\
&& + {2 \alpha^* \over m \kappa} 
\int d^2x d^2z  f^*(x)\, \bz \bdel_\bz f (z) \, S(z, x)
\label{df7}
\eeqar
Using (\ref{df6}) and (\ref{df7}) in (\ref{df5}), we find
\beqar
\bra{f} H \ket{f} &=& \bra{\nu} J^0(f^*) J^0(f) H \ket{\nu} + {2\over m} \int d^2x\, \del f^* \, \bdel f  \langle
\psi^\dagger \psi (x) \rangle \nonumber\\
&&+  {2  \over m \kappa} 
\int d^2x d^2z  \left[ \alpha^*f^*(x)\, \bz \bdel_\bz f (z) -
\alpha f^*(x) z\del_z f(z) \right] S(x,z)
\label{df8}
\eeqar
Here we have also used the fact that
$S(x, z) = S(z, x)$.
In evaluating this, we could have moved $H$ to the left end as well which will give the conjugate of the expression as written in (\ref{df8}).
Taking the average, we can write $\bra{f} H \ket{f} $ in a symmetrical way as
\beqar
\bra{f} H \ket{f}  &=&{1\over 2}\left[
\bra{\nu} J^0(f^*) J^0(f) H \ket{\nu} + \bra{\nu} H\,J^0(f^*) J^0(f) \ket{\nu} \right]
\nonumber\\
&&+ {2\over m} \int d^2x\, \del f^* \, \bdel f  \langle
\psi^\dagger \psi (x) \rangle \nonumber\\
&&+  {1  \over m \kappa} 
\int d^2x d^2z \Bigl[  f^*(x) \left (\alpha^*\, \bz \bdel_\bz f  -
\alpha  z\del_z f) (z) \right)\nonumber\\
&&\hskip .5in
+ f (x) \left (\alpha\, z \del_z f^* -
\alpha^*  \bz \bdel_\bz f^*) (z) \right) \Bigr] S(x, z)
\label{df9}
\eeqar
The action in (\ref{df2a}) is thus
\beqar
\S[f] &=&  \int d^2x d^2y\, {i \over 2} \left[ f^*(x) {\del f(y) \over \del t} - 
{\del f^*(x) \over \del t} f(y) \right] S(x, y)\nonumber\\
&&- {2\over m} \int d^2x\, \del f^* \, \bdel f  \langle
\psi^\dagger \psi (x) \rangle\nonumber\\
&&-  {1  \over m \kappa} 
\int d^2x d^2z \Bigl[  f^*(x) \left (\alpha^*\, \bz \bdel_\bz f  -
\alpha  z\del_z f) (z) \right)\nonumber\\
&&\hskip .5in
+ f (x) \left (\alpha\, z \del_z f^* -
\alpha^*  \bz \bdel_\bz f^*) (z) \right) \Bigr] S(x, z)
\label{df10}
\eeqar
In the last set of terms, we now do an integration-by-parts to move the derivatives on to $S(x, z)$. Further, we exchange $z\leftrightarrow x$ for some terms
to bring the argument of $f^*$ to be $x$ in all terms.
The result is 
\beqar
\S[f] &=&  \int d^2x d^2y\, {i \over 2} \left[ f^*(x) {\del f(y) \over \del t} - 
{\del f^*(x) \over \del t} f(y) \right] S(x, y)\nonumber\\
&&- {2\over m} \int d^2x\, \del f^* \, \bdel f  \langle
\psi^\dagger \psi (x) \rangle\nonumber\\
&&-  {1  \over m \kappa} 
\int d^2x d^2z \, f^*(x) f(z) \Bigl[ \alpha (z\del_z - x \del_x )
- \alpha^* (\bz \bdel_\bz - {\bar x} \bdel_{\bar x} )  \Bigr] S(x, z)
\label{df11}
\eeqar

Consider the evaluation of $S(x, z)$ using the wave functions
(\ref{erm16}). The phase factors involving $\Phi$ cancel out
and the factor $ \prod_{i<j} (\xi_i - \xi_j )^{2p+1}$ has identical scaling behavior for each of the $\xi$s. As a result
$z\del_z - x \del_x$ will vanish on such factors. Only the contribution
from the Gaussian exponential factors will be nonzero. Thus
\beq
(z\del_z - x \del_x ) S(x, z) = (\bz \bdel_\bz - {\bar x}  \bdel_{\bar x} ) 
S(x, z) = {1\over \kappa} ({\bar x} x - \bz z) S(x, z)
\label{df12}
\eeq
The action in (\ref{df11}) can thus be rewritten as
\beqar
\S[f] &=&  \int d^2x d^2y\, {i \over 2} \left[ f^*(x) {\del f(y) \over \del t} - 
{\del f^*(x) \over \del t} f(y) \right] S(x, y)\nonumber\\
&&- {2\over m} \int d^2x\, \del f^* \, \bdel f  \langle
\psi^\dagger \psi (x) \rangle\nonumber\\
&&-  {i {\dot \kappa}  \over 2 \kappa^2} 
\int d^2x d^2z \, f^*(x) f(z) ({\bar x} x - \bz z)  S(x, z)
\label{df11a}
\eeqar
where we used the expression for $\alpha$ from (\ref{erm25}).
This is the action in (\ref{df11b}) in text and used for deriving the equations of motion.
\section*{Appendix B: The action for the edge modes}
\def\theequation{B\arabic{equation}}
\setcounter{equation}{0}
In this Appendix, we give some of the details of the simplification of
the action (\ref{e31}).
The key point to keep in mind is that in taking the time-derivative of 
${\hat F}$, we get
\beqar
{\del \over \del t} {\hat F} &=& {\del \over \del t} \, e^{{\bar Z}\del_\bw}
e^{Z \del_w } \, F(\bw, w)\big\vert_{w=0}\nonumber\\
&=&  \Bigl[e^{{\bar Z}\del_\bw} {\dot{\bar Z}}
e^{Z \del_w } \, \del_\bw F(\bw, w) +
 e^{{\bar Z}\del_\bw}
e^{Z \del_w }  {\dot{Z}} \,\del_w F(\bw, w)\nonumber\\
&&\hskip .3in +
 e^{{\bar Z}\del_\bw}
e^{Z \del_w }   \, {\del F(\bw, w) \over \del t}\Bigr]_{w=0}
\label{A32}
\eeqar
There are terms involving the time-derivatives of $Z$, ${\bar Z}$.
From the expressions for $Z$ and ${\bar Z}$ in 
(\ref{e13}),
\beqar
{\dot Z} &=&{\del \over \del t} \left( e^{i \Phi}( z - \kappa D_\bz ) e^{-i \Phi} \right)
 =- {\dot \kappa} {\del\over \del \bz} + i {m {\ddot \kappa}\over 4}\, z
  \nonumber\\
{\dot{\bar Z}} &=& {\del \over \del t} \left( e^{i \Phi}( \bz + \kappa D_z ) e^{-i \Phi} \right)
= {\dot \kappa} {\del\over \del z} - i {m {\ddot \kappa}\over 4}\, \bz
\label{A33}
\eeqar
These also give us the relations
\beqar
&[Z, {\dot Z}] &= 0, \hskip .3in [{\bar Z}, {\dot{\bar Z}}] = 0
\nonumber\\
 &[Z, {\dot {\bar Z}}] &= - \eta, \hskip .2in
 [{\bar Z}, {\dot Z}] = \eta^*
  \label{A34}
 \eeqar
 where
 \beq
\eta = {{\dot \kappa} \over 2} + i {m\over 4} ({\dot \kappa}^2 - \kappa
 {\ddot\kappa} )
 \label{A34a}
 \eeq
For the simplification of the ${\hat F} {\dot{\hat F}}$-term in (\ref{e31}), 
we can use these results and write
\beqar
{i \over 2} \int \Psi^* {\hat F} {\dot{\hat F}} \Psi
&=& {i \over 2} \int \Psi^*
e^{{\bar Z} \del_\bu } e^{Z \del_u} F(\bu,u) 
\Bigl[e^{{\bar Z}\del_\bw} {\dot{\bar Z}}
e^{Z \del_w } \, \del_\bw F  +
 e^{{\bar Z}\del_\bw}
e^{Z \del_w }  {\dot{Z}} \,\del_w F \nonumber\\
&&\hskip .5in +
 e^{{\bar Z}\del_\bw}
e^{Z \del_w }   \, {\del F \over \del t}\Bigr] \Psi\nonumber\\
&=&{i \over 2} \int \Psi^*
e^{{\bar Z} (\del_\bw +\del_\bu) } e^{-\kappa \del_u \del_\bw} 
\left[ e^{Z \del_u} {\dot{\bar Z}}
e^{Z \del_w} \del_\bw F\, F(\bu,u)\right.\nonumber\\
&&\hskip .2in \left.
+ e^{Z (\del_u + \del_w)} {\dot Z} \del_w F\, F(\bu, u)
+ e^{Z (\del_u + \del_w)} (F {\dot F} ) \right] \Psi
\label{A35}
\eeqar
Although not indicated, it is implicit that we set $w = 0$, $u = 0$ at the end.
In the last term, we can act on $\Psi$ using
$Z \Psi = z\Psi$, so that we get the $*$-product of $F$ and ${\dot F}$.
In the second term, we use
\beq
{\dot Z} \Psi = (- {\dot \kappa} \del_\bz + i m {\ddot \kappa}/4)\,
e^{i \Phi - \bz z/2 \kappa} \, h(z)
= {z \over \kappa} \eta^* \Psi
\label{A36}
\eeq
So the second term in (\ref{A35}) can be written in terms of $\Psi^*(F * \del_z F) \eta^* z/\kappa \, \Psi$. For the first term, we have to move
$e^{Z \del_u}$ to the right so that it can act on $\Psi$ to
give $e^{z \del_u}$. This is done by the identity
\beq
e^{Z \del_u} {\dot {\bar Z}} = e^{Z \del_u} {\dot {\bar Z}} e^{-Z \del_u} 
\, e^{Z \del_u} =
({\dot{\bar Z}} - \eta \del_u ) e^{Z \del_u} 
\label{A37}
\eeq
where we used (\ref{A34}). Further, since ${\dot{\bar Z}}$ commutes with
${\bar Z}$, we can move it to the left end and evaluate it
by integration by parts. This gives
\beqar
\int \Psi^* {\dot {\bar Z}} \chi &=&
\int \Psi^* ( {\dot \kappa} \del_z - i m {\ddot \kappa} \bz /4) \chi
= \int \left[ -{\dot \kappa} \del_z (-i \vf - \bz z/2 \kappa) \Psi^* \chi\right]
- i {m {\ddot \kappa}\over 4} \Psi^* \bz \chi\nonumber\\
&=&\int {\bz \over \kappa} \eta\, \Psi^* \chi
\label{A38}
\eeqar
The first term on the right hand side of (\ref{A35}) can thus be written
in terms of $- \eta \del_zF * \del_\bz F + (\bz /\kappa) \eta
F*\del_\bz F$. Combining all terms
\beqar
{i \over 2} \int \Psi^* {\hat F} {\dot{\hat F}} \Psi
&=& {i \over 2} \int \Psi^* 
\biggl[ \left(F* {\del F \over \del t}\right) + \eta {\bz \over \kappa} (F* \del_\bz F) + \eta^* {z\over \kappa} (F* \del_z F)\nonumber\\
&&\hskip .6in - \eta (\del_z F* \del_\bz F)
\biggr] \Psi
\label{A39}
\eeqar
Now we consider the term with the other ordering in (\ref{e31}).
\beqar
{i \over 2} \int \Psi^*  {\dot{\hat F}} {\hat F} \Psi
&=& {i \over 2} \int \Psi^* \biggl[
e^{{\bar Z} \del_\bw } e^{Z \del_w} {\del F \over \del t} 
+ {\dot{\bar Z}} e^{{\bar Z} \del_\bw } e^{Z \del_w} \del_\bw F
\nonumber\\
&&\hskip .5in 
+ e^{{\bar Z} \del_\bw } e^{Z \del_w} {\dot Z}  \del_w F
\biggr] e^{{\bar Z} \del_\bu } e^{Z \del_u} F(\bu, u) \Psi
\nonumber\\
&=& {i \over 2} \int \Psi^* \biggl[ \left( {\del F \over \del t}* F \right)
+ \eta {\bz \over \kappa} (\del_\bz F * F ) \biggr] \Psi\nonumber\\
&&+ {i \over 2} \int \Psi^* 
e^{{\bar Z} \del_\bw } e^{Z \del_w} \del_w F \, {\dot Z} 
e^{{\bar Z} \del_\bu } e^{Z \del_u} F\Psi
\label{A40}
\eeqar
We move ${\dot Z}$ to the right to act on $\Psi$ as
follows.
\beqar
{\dot Z} e^{{\bar Z} \del_\bu } e^{Z \del_u} F\Psi&=&
e^{{\bar Z} \del_\bu }  \left( [{\dot Z}, {\bar Z}] \del_\bu 
+ {\dot Z} \right) e^{Z \del_u} F\Psi\nonumber\\
&=& e^{{\bar Z} \del_\bu } \left[ (- \eta^* \del_\bu ) e^{Z \del_u} F
+ e^{Z \del_u} F {\dot Z} \right]\Psi\nonumber\\
&=& e^{{\bar Z} \del_\bu }  e^{Z \del_u}
\left[ - \eta^* \del_\bu F + {z \over\kappa}\eta^* F \right] \Psi
\label{A41}
\eeqar
Using this back in (\ref{A40}), we get the result
\beqar
{i \over 2} \int \Psi^*  {\dot{\hat F}} {\hat F} \Psi
&=& {i \over 2} \int \Psi^* \biggl[ \left( {\del F \over \del t}* F \right)
+ \eta {\bz \over \kappa} (\del_\bz F * F ) \nonumber\\
&&\hskip .2in
+\eta^* {z\over \kappa} (\del_z F*F) - \eta^* (\del_z F* \del_\bz F)
\biggr] \Psi
\label{A42}
\eeqar
Combining this with (\ref{A39}), the action (\ref{e31}) takes the form
\beqar
\S&\approx& {i \over 2} \int \Psi^* \biggl[
\left( F*{\del F \over \del t}  - {\del F \over \del t}* F \right)
+ (\eta^* - \eta ) (\del_z F* \del_\bz F)\nonumber\\
&&\hskip .4in 
+\eta {\bz \over \kappa} \left(  (F* \del_\bz F ) - (\del_\bz F*F )\right)
+ \eta^* {z \over \kappa} \left(  (F* \del_z F ) - (\del_z F*F )\right)
\biggr] \Psi\nonumber\\
&&\hskip .4in
+ {1\over 2} \int \Psi^* [{\hat F} , [{\hat F}, H']] \Psi
\label{A43}
\eeqar

We now need to address the simplification of the double commutator in the last term of this expression.
Since we are considering the integer Hall effect, the
Hamiltonian has the form
$H' = H_0 +V$, where we take the confining potential $V$ to be an operator
made of $Z$ and ${\bar Z}$. 
(The interparticle interaction terms are expected to be not very significant.)
The products involving $V$
then reduce to
$*$-products as in (\ref{e18}) and we can write
\beq
{1\over 2} \int \Psi^* [{\hat F} , [{\hat F}, V]] \Psi
= {1\over 2} \int \Psi^* \left[ F* (F* V - V*F) - (F* V - V* F )*F
\right] \Psi
\label{e44}
\eeq

The terms involving operator $H_0$ are a bit trickier since
it cannot be written entirely in terms of
$Z$ and ${\bar Z}$. Its reduction is therefore more involved.
For the terms involving commutators with $H_0$, we note that using
the formula (\ref{erm6}) for $H_0$, we get the following relations:
\beqar
{}[H_0, {\bar Z}] &=& \left( - {2 / m}\right) (\del_z - \lambda \bz )
[ \del_\bz + \lambda z, {\bar Z}]\nonumber\\
&=& \left( 2/ m\right) \alpha (\del_z - \lambda \bz )
\nonumber\\
{}[H_0, Z] &=& - \left( 2/ m\right) [ \del_z - \lambda \bz , Z] (\del_\bz + \lambda z )\nonumber\\
&=& \left( 2/ m\right) \alpha^* (\del_\bz + \lambda z)
\nonumber\\
{} [H_0, {\bar Z}], {\bar Z}] &=& 0,
\hskip .3in
 [H_0, {Z}], { Z}]  = 0
 \label{A45}
 \eeqar
 where $\alpha$ is as given in (\ref{erm25}).
We then find
\beqar
[{\hat F}, H_0] &=& e^{{\bar Z}\del_\bw } e^{Z \del_w} F\, H_0
- H_0 \, e^{{\bar Z}\del_\bw } e^{Z \del_w} F\nonumber\\
&=&e^{{\bar Z}\del_\bw } e^{Z \del_w}  \left[
F H_0 - e^{-Z \del_w} e^{-{\bar Z}\del_\bw } H_0 \,e^{{\bar Z}\del_\bw } e^{Z \del_w}  F \right]\nonumber\\
&=&e^{{\bar Z}\del_\bw } e^{Z \del_w}  \left[
F H_0 - e^{-Z \del_w} \left\{
 H_0 + {2\over m} \alpha (\del_z - \lambda \bz ) \del_\bw
 \right\} e^{Z \del_w}  F \right]\nonumber\\
 &=&e^{{\bar Z}\del_\bw } e^{Z \del_w}  \biggl[
F H_0 - e^{-Z \del_w} \left\{
 H_0 + {2\over m} \alpha (\del_z - \lambda \bz ) \del_\bw
 + {2\over m} \alpha^* \del_w (\del_\bz + \lambda z)\right.\nonumber\\
 &&\hskip .7in \left.
 - {2\over m} \alpha \alpha^* \del_\bw \del_w 
 \right\} \, F \biggr]\nonumber\\
 &=&- {2\over m} 
 e^{{\bar Z}\del_\bw } e^{Z \del_w}  \biggl[ \alpha^* \,\del_w F \, (\del_\bz + \lambda z )+
 \alpha\, \del_\bw F \, (\del_z - \lambda \bz )   - \alpha^* \alpha\, \del_\bw \del_w F \biggr]
 \label{A47}
\eeqar
Our next step is to simplify ${\hat F} [{\hat F}, H_0]$. We write this out as
\beqar
\int \Psi^* {\hat F} [{\hat F}, H_0] \Psi
&=& {\rm Term~1} + {\rm Term~2} + {\rm Term~3}\nonumber\\
{\rm Term~1}&=&- {2\over m}  \int \Psi^* \,e^{{\bar Z}\del_\bu } e^{Z \del_u} F(\bu, u)
 e^{{\bar Z}\del_\bw } e^{Z \del_w}  \Bigl[
 \alpha^* \,\del_w F \, (\del_\bz + \lambda z ) \Bigr] \Psi\nonumber\\
 {\rm Term ~2}&=&- {2\over m}  \int \Psi^* \,e^{{\bar Z}\del_\bu } e^{Z \del_u} F(\bu, u)
 e^{{\bar Z}\del_\bw } e^{Z \del_w} \Bigl[
  \alpha\, \del_\bw F \, (\del_z - \lambda \bz ) \Bigr] \Psi
  \nonumber\\
  {\rm Term~3}&=&- {2\over m}  \int \Psi^* \,e^{{\bar Z}\del_\bu } e^{Z \del_u} F(\bu, u)
 e^{{\bar Z}\del_\bw } e^{Z \del_w} \Bigl[- \alpha^* \alpha\, \del_\bw \del_w F \Bigr] \Psi
 \label{A48}
\eeqar
In ${\rm Term~1}$, we move the ${\bar Z}$-terms to the left, and since
$Z$ commutes with $(\del_\bz + \lambda z)$, it can then act on
$\Psi$ and give $z$. Thus
\beq
{\rm Term~1}= - {2\over m} \int \Psi^* \,\alpha^* 
\bigl( F* \del_z F \bigr)  (\del_\bz + \lambda z) \Psi
\label{A49}
\eeq
For ${\rm Term~2}$, since $Z$ does not commute with
$(\del_z - \lambda \bz)$, we use
\beqar
e^{Z (\del_u + \del_w) } F(\bu, u) \del_\bw F (\del_z - \lambda \bz ) \Psi
&=& e^{Z (\del_u + \del_w) } (\del_z - \lambda \bz ) \Psi
\, F(\bu, u) \del_\bw F \nonumber\\
&=& \left[(\del_z - \lambda \bz ) \, e^{Z (\del_u + \del_w) } \Psi
+ \alpha^* (\del_u + \del_w)  e^{Z (\del_u + \del_w) } \Psi
\right] \nonumber\\
&&\hskip .7in \times F(\bu, u) \del_\bw F \nonumber\\
&=& e^{z (\del_u +\del_w) } \Bigl[
(1+ \alpha^*) (\del_u +\del_w ) F\, \del_\bw F\nonumber\\
&&\hskip .7in
+ F\, \del_\bw F (\del_z - \lambda \bz ) \Bigr] \Psi
\label{A50}
\eeqar
With this result,
\beqar
{\rm Term~2} &=&- {2\over m} \int \Psi^*
\biggl[ \alpha (1+ \alpha^*) \left( (\del_z F* \del_\bz F )
+ ( F* \del_z \del_\bz F) \right) \nonumber\\
&&\hskip .5in + \alpha
(F*\del_\bz F) (\del_z - \lambda \bz ) \biggr] \Psi
\label{A51}
\eeqar 
The last term is straightforward.
\beq
{\rm Term ~3}= - {2\over m} \int \Psi^* 
\left[ - \alpha \alpha^* ( F* \del_z \del_\bz F) \right] \Psi
\label{A52}
\eeq
Collecting results,
\beqar
\int \Psi^* {\hat F} [{\hat F}, H_0] \Psi
&=& - {2\over m} \int \Psi^*
\biggl[ \alpha^* (F* \del_z F) (\del_\bz + \lambda z)
+ \alpha (F*\del_\bz F) (\del_z - \lambda \bz )\nonumber\\
&&\hskip .7in  + \alpha (1+ \alpha^*)
(\del_z F* \del_\bz F ) + \alpha (F* \del_z \del_\bz F) \biggr] \Psi
\label{A53}
\eeqar
In a similar way, we can calculate $[{\hat F}, H_0] {\hat F}$.
\beqar
\int \Psi^*  [{\hat F}, H_0] {\hat F} \Psi
&=& {\rm Term ~4} + {\rm Term~5} + {\rm Term~6}\nonumber\\
{\rm Term ~4}  &=& - {2\over m} \int \Psi^* \,  e^{{\bar Z}\del_\bw } e^{Z \del_w}
 \alpha^* \del_w F (\del_\bz + \lambda z)e^{{\bar Z}\del_\bu } e^{Z \del_u} F(\bu, u)  \Psi\nonumber\\
 {\rm Term~5}&=&
  - {2\over m} \int \Psi^* \,  e^{{\bar Z}\del_\bw } e^{Z \del_w}
\alpha \del_\bw F (\del_z - \lambda \bz) e^{{\bar Z}\del_\bu } e^{Z \del_u} F(\bu, u) \Psi \nonumber\\
{\rm Terr~6}&=&   - {2\over m} \int \Psi^* \,  e^{{\bar Z}\del_\bw } e^{Z \del_w}\Bigl[
- \alpha \alpha^* 
\del_\bw \del_w F \Bigr] e^{{\bar Z}\del_\bu } e^{Z \del_u} F(\bu, u) \Psi
\label{A54}
\eeqar
We can now use the identity
\beq
(\del_\bz + \lambda z )\, e^{ {\bar Z}\del_\bu}
= e^{ {\bar Z}\del_\bu}\,  \left( \del_\bz + \lambda z + \alpha \del_\bu\right)
\label{A55}
\eeq
This simplifies ${\rm Term~4}$ and ${\rm Term~5}$ as
\beqar
{\rm Term~4}&=&  - {2\over m} \int \Psi^* \,  e^{{\bar Z}\del_\bw } e^{Z \del_w} e^{{\bar Z}\del_\bu} \alpha^* \del_w F
\left( \del_\bz + \lambda z + \alpha \del_\bu \right) e^{Z \del_u} F \Psi
\nonumber\\
&=& - {2\over m} \int \Psi^* \,  e^{{\bar Z}(\del_\bw +\del_\bu)} 
e^{ -\kappa \del_\bu \del_w} \, \alpha^* 
\left( \del_\bz + \lambda z + \alpha \del_\bu \right) e^{Z (\del_u +\del_w)} \del_w F\, F \Psi\nonumber\\
&=&- {2\over m} \int \Psi^* \,  e^{{\bar Z}(\del_\bw +\del_\bu)} 
e^{ -\kappa \del_\bu \del_w} \, \alpha^*  
e^{z (\del_u +\del_w)} \Bigl[
(\del_\bz + \lambda z ) \del_w F\, F + \alpha \del_w F \, \del_\bu F \Bigr] \Psi\nonumber\\
&=& - {2\over m} \alpha^* \int \Psi^* \Bigl[
(\del_z F* F)  (\del_\bz + \lambda z) + \alpha (\del_z F* \del_\bz F)
\Bigr] \Psi
\label{A56}
\eeqar
\beqar
{\rm Term~5}&=&\!\!\!- {2\over m} \alpha \int \Psi^*
e^{{\bar Z}\del_\bw } e^{Z \del_w} \del_\bw F 
e^{{\bar Z} \del_\bu} (\del_z - \lambda \bz ) e^{z \del_u} F \Psi
\nonumber\\
&=&\!\!\!- {2\over m} \alpha \int \Psi^*
e^{{\bar Z}(\del_\bw + \del_\bu) } e^{-\kappa \del_w \del_\bu} e^{Z \del_w} \del_\bw F 
e^{z \del_u} (\del_z - \lambda \bz + \del_u) F \Psi
\nonumber\\
&=&\!\!\!- {2\over m} \alpha \int \Psi^*
e^{{\bar Z}(\del_\bw + \del_\bu) } e^{-\kappa \del_w \del_\bu} 
\biggl[ e^{z (\del_u +\del_w) }\del_\bw F \del_u F
+ e^{Z \del_u} \del_\bw F e^{z \del_u} F (\del_z - \lambda \bz) \biggr] \Psi
\nonumber\\
&\!=&\!\!\!- {2\over m} \alpha \int \Psi^*
\biggl[ (\del_\bz F * \del_z F) + (1 + \alpha^*) (\del_z\del_\bz F*F)
+ (\del_\bz F* F) (\del_z - \lambda \bz ) \biggr] \Psi
\nonumber\\
\label{A57}
\eeqar
\beq
{\rm Term~6} = -{2\over m} \int \Psi^* \left[ - \alpha \alpha^* (\del_z \del_\bz F*F) \right] \Psi
\label{A58}
\eeq
Again, collecting results,
\beqar
\int \Psi^*  [{\hat F}, H_0] {\hat F} \Psi
&=& -{2\over m} \int \Psi^* \biggl[ 
\alpha^* (\del_z F*F) (\del_\bz + \lambda z) 
+ \alpha^* \alpha (\del_z F * \del_\bz F) \nonumber\\
&&\hskip .7in
+ \alpha (\del_\bz F *F) (\del_z - \lambda \bz)
+ \alpha (\del_\bz F* \del_z F) \nonumber\\
&&\hskip .7in + \alpha (\del_z \del_\bz F*F)
\biggr] \Psi
\label{A59}
\eeqar
Now we can combine this with (\ref{A53}) to write
\beqar
\int \Psi^* [{\hat F},  [{\hat F}, H_0] ] \Psi
&=&-{2\over m} \int \Psi^* \biggl[ 
\alpha^* \left( (F*\del_z F) - (\del_z F*F) \right) (\del_\bz + \lambda z)\nonumber\\
&&\hskip .7in
+ \alpha \left( (F*\del_\bz F) - (\del_\bz F*F) \right) (\del_z - \lambda \bz)
\nonumber\\
&&\hskip .7in
+ \alpha \left( (\del_z F* \del_\bz F) - (\del_\bz F* \del_z F) \right) 
\nonumber\\
&&\hskip .7in
+ \alpha \left( (F* \del_z \del_\bz F) - (\del_z \del_\bz F* F) \right) \biggr] \Psi
\label{A60}
\eeqar
There are some apparent asymmetries in this expression.
This is because we have kept both the derivative operators as acting to the right on $\Psi$.
Consider taking the complex conjugate of the first term on the right hand side of (\ref{A60}). Writing $f = (F*\del_z F - \del_z F*F )$ for brevity,
\beqar
\left\{\int \Psi^* \alpha^* \,f\, (\del_\bz + \lambda z) \Psi
\right\}^*
&=&\alpha \int \left( \del_z \Psi^* + \lambda \bz  \Psi^*\right) 
f^* \Psi\nonumber\\
 &=& -\alpha \int \left[ \Psi^* ( \del_z f^* ) \Psi
 + \Psi^*  f^* (\del_z - \lambda \bz ) \Psi \right]\nonumber\\
 &=& \int \Psi^* \biggl[
 \alpha ( F*\del_\bz F - \del_\bz F*F) (\del_z - \lambda \bz ) \nonumber\\
 &&\hskip .5in
 + \alpha (F* \del_z \del_\bz F - \del_z \del_\bz F* F )\nonumber\\
 &&\hskip .5in + \alpha
 (\del_z F* \del_\bz F - \del_\bz F * \del_z F) \biggr] \Psi
 \label{A61}
\eeqar
These are the last three terms in (\ref{A60}). Thus we can write
(\ref{A60}) as
\beq
\int \Psi^* [{\hat F},  [{\hat F}, H_0] ] \Psi
=-{2\over m} \int \Psi^* \Bigl[ 
\alpha^* \left( (F*\del_z F) - (\del_z F*F) \right) (\del_\bz + \lambda z)\Bigr] \Psi
+ {\rm c.c.}
\label{A62}
\eeq
where $\alpha$ is as given in (\ref{erm25}).
Using (\ref{A42}), (\ref{e44}) and (\ref{A62}), we  can write the action
to quadratic order in $F$ as
\beqar
\S&\approx& {i \over 2} \int \Psi^* \biggl[
\left( F*{\del F \over \del t}  - {\del F \over \del t}* F \right)
+ (\eta^* - \eta ) (\del_z F* \del_\bz F)\nonumber\\
&&\hskip .4in 
+\eta {\bz \over \kappa} \left(  (F* \del_\bz F ) - (\del_\bz F*F )\right)
+ \eta^* {z \over \kappa} \left(  (F* \del_z F ) - (\del_z F*F )\right)
\biggr] \Psi\nonumber\\
&&\hskip .4in
+ {1\over 2} \int \Psi^* \left[ F* (F* V - V*F) - (F* V - V* F )*F
\right] \Psi\nonumber\\
&&\hskip .4in
- {1\over m} 
\int \left\{ \Psi^* \Bigl[ 
\alpha^* \left( (F*\del_z F) - (\del_z F*F) \right) (\del_\bz + \lambda z)\Bigr] \Psi
+ {\rm c.c.}\right\}
\label{e63}
\eeqar
It is useful to simplify this by evaluating the $*$-commutators to the lowest
nontrivial order, in terms of Poisson brackets. The relevant expansions are
\beqar
\eta {\bz \over \kappa} (F* \del_\bz F - \del_\bz F* F)
&=& - \eta \bz  \left( \del_z F \,\del_\bz \del_\bz F -
\del_z \del_\bz F \del_\bz F \right) + \cdots\nonumber\\
&=& -\eta \left[ \del_z F (\del_\bz (\bz \del_\bz F) -
\del_z (\bz \del_\bz F) \del_\bz F - \del_z F \del_\bz F \right] + \cdots
\nonumber\\
&=& - i \eta \{ F, \bz \del_\bz F\} + \eta \del_z F \del_\bz F + \cdots
\label{e64}\\
\eta^* {z \over \kappa} (F* \del_z F - \del_z F* F)
&=& -i \eta^* \{ F, z \del_z F\} - \eta^* \del_z F \del_\bz F + \cdots
\label{e65}\\
F*{\del F \over \del t}  - {\del F \over \del t}* F
&=&- i \kappa \{F, {\del F \over \del t} \} + \cdots\nonumber\\
F*\del_z F - \del_z F *F &=&
- i \kappa \{ F, \del_z F\} +\cdots
\label{e66}
\eeqar
Combining all these results
\beqar
\S &=& \int \Psi^* \biggl[  {\kappa \over 2} \{F, {\dot F}\} + {\eta \over 2}
\{ F, \bz \del_\bz F\} + {\eta^* \over 2} \{F, z\del_z F\} \biggr] \Psi
\nonumber\\
&& \hskip .5in
+ {\kappa\over m} \biggl[ i\int \Psi^* \alpha^* \{F, \del_z F\} (\del_\bz + \lambda z) \Psi + c.c.\biggr]\nonumber\\
&&\hskip .5in
-{\kappa^2\over 2} \int \Psi^* \{ F,\{F, V\} \} \Psi +\cdots
\label{e67}
\eeqar

A very useful point for further simplification is the following.
All the terms in the integrals for the action involve Poisson brackets.
Even the term with $(\del_\bz + \lambda z)$ acting on $\Psi$ arises
from the double commutator in (\ref{A62}). Therefore even though
$F = \sum_i F(z_i, \bz_i)$, there is no mixing between 
the coordinates of different particles in the
integrand, except in $\Psi^*$ and $\Psi$.
Therefore, by virtue of the permutation symmetry of
the $\Psi$'s, we can use the result
\beq
\int \Psi^* \, \sum_i {\cal O }(z_i, \bz_i ) \Psi = N \int \Psi^* \, {\cal O }(z_1, \bz_1) \Psi
\label{e67a}
\eeq

We also note that, since the integration is carried out with the phase volume, we have
\beq
\int \{ F, G\} = 0
\label{e68}
\eeq
We can use this relation to transfer some of the Poisson brackets to
$J^0 = \Psi^* \Psi$. 
Thus
\beqar
\S &=& N \Biggl[\int \{F, J^0 \} \left[ -{\kappa \over 2} {\dot F} - {\eta\over 2}
\bz \del_\bz F - {\eta^* \over 2} z\del_z F  + {\kappa^2\over 2} \{F, V\}\right]\nonumber\\
&&\hskip .2in
+ {\kappa\over m} \biggl[ i\int \Psi^* \alpha^* \{F, \del_z F\} (\del_\bz + \lambda z) \Psi + c.c.\biggr]\Biggr]
\label{e69}
\eeqar
The terms $-{\half}\{F, J^0\}  \kappa {\dot F}$ and
$-{\half} \{F, J^0\} \{ F, V\}$ are the terms we expect for the edge dynamics in the absence of any time-dependence for the magnetic field.
From the definition of $\alpha$ in (\ref{erm25}) and
$\eta$ in (\ref{A34a}), we see that that they vanish
if the field is independent of time, with $\lambda = \lambda_0$.
We see that, as a consistency check, all the terms in (\ref{e69})
which are extra to what is the usual edge state dynamics
will vanish for $\lambda = \lambda_0$.

The last term in (\ref{e69}) can be simplified further.
We use $(\del_\bz + \lambda z )\Psi = (\alpha z /\kappa)\Psi$ and write out the Poisson brackets to bring it to the form
\beqar
{\kappa\over m} \biggl[ i\int \Psi^* \alpha^* \{F, \del_z F\} (\del_\bz + \lambda z) \Psi + c.c.\biggr] &=&
- i {\alpha^*\alpha \over m}\int \{ F, J^0\} (z\del_z - \bz \del_\bz ) F\nonumber\\
&&+ {2 \alpha^* \alpha \over m} \int J^0\, \del F \bdel F
\label{e70}
\eeqar
The terms with $\{F, J^0\}$ will reduce to a term on the  boundary of the droplet.
 The last term in this equation is a bulk term. 
 To simplify things further, consider a spherical droplet. The radius $R$ of the droplet is given, for $\nu = 1$, by
 \beq
 R^2 = \kappa (N-1) \approx N \kappa 
 \label{e71}
 \eeq
We also have uniform density in the large $N$ limit, of the form
\beq
J^0 = {1 \over \pi R^2}\, \Theta (R^2 - \bz z)
\label{e72}
\eeq
where $\Theta$ is the step function. This leads to
\beq
\{ F, J^0\} =  {i \over \pi R^2} (z \del_z - \bz \del_\bz ) F \, \delta (\bz z- R^2)
=  {1 \over \pi R^2} \delta (\bz z- R^2) \,{\del F \over \del \theta} 
\label{e73}
\eeq
where we used the parametrization $z = r e^{i \theta}$.
The action (\ref{e69}) now reduces to
\beqar
\S &=& N\Biggl[ \int dt \oint {d \theta \over 2 \pi R^2} 
\biggl[- {\kappa \over 2} {\dot F} {\del F \over \del \theta} -
{\eta + \eta^* \over 4} \left( r {\del F \over \del r} {\del F \over \del \theta}
\right) \nonumber\\
&&\hskip .2in - \left( {\alpha \alpha^* \over m}  + i {(\eta - \eta^* )\over 4}\right) {\del F \over \del \theta}
{\del F \over \del \theta} -{\kappa^2\over 2} \left( {\del F \over \del \theta}\right)^2 {\del V \over \del r^2}\biggr]\nonumber\\
&&+ \int dt d^2x  {\alpha^* \alpha \over 2\pi R^2 m} \Theta(R^2 - r^2) 
(\nabla F )^2\Biggr]
\label{e74}
\eeqar 
From (\ref{A34a}), $\eta+ \eta^* = {\dot \kappa}$.
If we consider $F$ as a function of $t, R$ and $\theta$ on the edge,
\beq
- {\kappa \over 2} {d F \over d t} = - {\kappa \over 2} \left[{\del F \over \del t} + {{\dot R}\over R}
 \,R {\del F \over \del R} \right] = - {\kappa \over 2}  {\del F \over \del t} - {{\dot \kappa}\over 4}
 R {\del F \over \del R}
 \label{e75}
 \eeq
 where we used the fact that $R \sim\sqrt{\kappa}$ from  
 (\ref{e71}). So the first two terms in the action (\ref{e74}) can be combined into a time-derivative taking account of the time-dependence of the radius of the
 droplet. By using the Ermakov equation (\ref{erm10})
 to rewrite ${\ddot{\kappa}}$ in the expression for $\eta$, we find
 \beq
 {\alpha \alpha^* \over m}  + i {(\eta - \eta^* )\over 4}
 =  {\kappa \over 2}
\,{2 \over m} \left[ \lambda^2 \left( \kappa - {1\over 2 \lambda_0}
 \right) - \left( \lambda - {1\over 2 \kappa}\right) \right] \equiv 
 {\kappa \over 2} \, \beta
 \label{e90}
 \eeq
 The action now takes the form
\beqar
\S &=& N\Biggl[\int dt \oint d \theta {\kappa\over 2 \pi R^2} 
\left[- {1 \over 2} {d F \over d t} {\del F \over \del \theta} 
- {1\over 2} \left( \kappa {\del V \over \del r^2} +\beta\right) {\del F \over \del \theta}{\del F \over \del \theta} \right]\nonumber\\
&&+ \int_D dt d^2x  {\alpha^* \alpha \over 2\pi R^2 m}
(\nabla F )^2\Biggr]\nonumber\\
&=&- {1\over 4\pi} \int dt \oint d \theta  
\left[ {d F \over d t} {\del F \over \del \theta}
+ \left( \kappa {\del V \over \del r^2} + \beta\right) {\del F \over \del \theta} {\del F \over \del \theta}\right]\nonumber\\
&&
+ \int_D dt d^2x  {\alpha^* \alpha \over 2\pi \kappa m}
(\nabla F)^2
\label{e91}
\eeqar 
where the subscript $D$ on the integral indicates that the regions of integration is a disc of radius $R$.
This is the action quoted in text in (\ref{e91a}) and used for the equations of motion.


\end{document}